\documentclass[aps,twocolumn, pre]{revtex4}

\usepackage{graphicx}
\def\al{\alpha}
\def\w{p}
\def\pc{\pi}
\def\T{S}

\def\a{\rho}
\def\n{\n}
\def\g{Q}
\def\s{t_0}

\def\I{n}
\def\t{t}
\def\zt{{\tilde Z}}
\def\ord{{\cal O}}
\def\AB{BA\ }

\begin{document}

\title{Parallel Dynamics and Computational Complexity of Network Growth Models}

\author{Benjamin Machta$^{1,2}$}
\author{Jonathan Machta$^1$}
\affiliation{$^1$Department of Physics, University of
Massachusetts, Amherst, MA 01003-3720}
\affiliation{$^{2}$Department of Physics, Brown University,
         Box 1843, Providence, RI 02912}
\begin{abstract}

The parallel computational complexity or depth of growing network models is investigated. The networks considered are generated by preferential attachment rules where the probability of attaching a new node to an existing node is given by a power, $\alpha$ of the connectivity of the existing node.  Algorithms for generating growing networks very quickly in parallel are described and studied.   The sublinear and superlinear cases require distinct algorithms. As a result, there is a discontinuous transition in the parallel complexity of sampling these networks corresponding to the discontinuous structural transition at $\alpha=1$, where the networks become scale free.  For $\alpha>1$ networks can be generated in constant time while for $0 \leq \alpha < 1$ logarithmic parallel time is required.  The results show that these networks have little depth and embody very little history dependence despite being defined by sequential growth rules.

\end{abstract}

\maketitle

\section{Introduction}

This paper is concerned with the complexity of networks.  Many features of biological, social and technological systems can be described in terms of networks.  Examples include gene networks, friendship networks, citation networks, the power grid, the internet and the world wide web~\cite{Ne00}.  Although the systems that generate these networks are extremely complex, the networks themselves may or may not evidence this complexity.   In many cases the networks generated by complex systems are approximately scale free.  Barabasi and Albert~\cite{BaAl99} (BA) showed that scale free networks can be generated by rules for network growth that embody the intuitively plausible idea of preferential attachment.  In their model, the network grows by the addition of one node at a time and each node creates one new connection to an existing node. Existing nodes in the network that already have many connections are more likely to gain the new connection from the new node added to the network.  The growing network model seems to incorporate a history dependent process, albeit simplified, into the generation of the network.

One of the essential markers of complexity is a long history.  Complex systems cannot arise instantaneously but require a long sequence of interactions to develop.  Neither ``complexity'' nor ``long history'' are well-defined concepts but an appropriate proxy for these ideas can be formulated within computational complexity theory.  Computational complexity theory is concerned with the resources required to solve problems.  Although there are various resources required to solve computational problems, here we focus on parallel time or depth.  Depth is the number of computational steps needed by a parallel computer to solve a problem. In our case, the problem is to generate a statistically correct representation of the network.  If the depth of the computation needed to generate the network is large, even using the most efficient algorithm, we say that the network has a long history and cannot be generated quickly.  If, on the other hand, only a few parallel steps are needed to generate the network, then it cannot be complex.

The \AB growing network model would appear to have substantial depth since nodes are added to the network one at a time and the preferential attachment rule uses knowledge of the existing state of the network to decide where each new node will attach.  If the \AB model captures the mechanism for the scale free behavior found in real world networks then perhaps one can conclude that some of the complexity or history dependence of the social, biological or technological system that generated the network is embodied in the network.  One of the main conclusions of this paper is that growing network models do not actually embody much history dependence.  What we show is that there is a fast parallel algorithm that generates \AB  growing networks with $N$ nodes in $\ord (\log \log N)$ steps.

The \AB model has a linear preferential attachment rule.  Krapivsky, Redner and  Leyvraz~\cite{KrReLe00} introduced a generalization of the \AB model in which the probability to connect to a node is proportional to a power, $\al$ of its number of connections. The original \AB model is the case $\al=1$ while $\al=0$ is a random network.  The class of models  $0 \leq \al < \infty$ is analyzed in  Refs.\ \cite{KrReLe00,KrRe00} and it is seen that $\al=1$ marks a ``phase transition'' between a ``high temperature phase" for $\al<1$ where no node has an extensive number of connections and a ``low temperature phase'' for $\al>1$ where a single node has almost all connections in the large $N$ limit.  

We show that distinct but related parallel algorithms are needed to efficiently simulate the $\al<1$ and $\al>1$ regimes so that there is a discontinuous transition in the computational complexity of simulating the model at $\al=1$. For $0<\al<1$ the parallel time for generating a network of size $N$ scales logarithmically in $N$ while for $1<\al<\infty$ there is a constant time algorithm. Exactly at $\al=1$ yet a third algorithm is most efficient with parallel running time that is $\ord(\log \log N)$.  

A number of non-equilibrium models in statistical physics defined by sequential rules have been shown to have fast parallel dynamics.  Examples include  the Eden model, invasion percolation, the restricted solid-on-solid model~\cite{MaGr}, the Bak-Sneppen model~\cite{MaLi01} and internal diffusion-limited aggregation~\cite{MoMa00} all of which can be simulated in parallel in polylogarithmic time. On the other hand, no polylog time algorithm is known for generating diffusion-limited aggregation clusters and there is evidence that only powerlaw speed-ups are possible using parallelism~\cite{MaGr96,TiMa04}.    

Phase transitions in computational complexity have been the object of considerable recent study, for example, see Ref.\ \cite{MoZeKiSeTr99}. Most of the attention has been focused on {\bf NP}-hard combinatorial optimization problems.  
Growing networks and many other physically motivated models are naturally related to problems in the lower class {\bf P} (problems solvable in polynomial time).  One of the purposes of this paper is to provide an example of a transition in computational complexity at this lower level of the complexity hierarchy.  
 
The paper is organized as follows.  In the next section we define and describe the class of preferential attachment network growth to be studied.  In Sec.\ \ref{sec:comp} we give a brief review of relevant features of parallel computational complexity theory.  Section \ref{sec:alg} presents efficient parallel algorithms for sampling growing network models and related systems, Sec.\ \ref{sec:eff} analyzes the efficiency of these algorithms and Sec.\ \ref{sec:sim} presents results from numerical studies of the efficiency of one of the algorithms.  The paper ends with a discussion.

\section{Growing Network Models}
\label{sec:growth}

	In this section we describe growing network models with preferential attachment first considered by Barabasi and Albert \cite{BaAl99} and later generalized by Krapivsky, Redner and Leyvraz~\cite{KrReLe00,KrRe00}.  
	Consider a graph with $N$ ordered nodes, each having one outgoing link, constructed by the addition of one node every time step so that at time $\t$ in the construction, node $\t$ is attached to a previous node, $0$ through $\t-1$.  The probability $\pc_{\I}(t)$ of attaching node $\t$ to node $\I<\t$ is given by 
\begin{equation}
\label{eq:pat}
\pc_{\I}(t) = \frac{F(k_{\I}(t))}{Z(t)}
\end{equation}
where $k_{\I}(t)$ is the degree (number of connections) of $\I$, at time t, $F$ is some function and $Z$ is the normalization given by
\begin{equation}
Z(t) = \sum_{j=0}^{t-1}F(k_{j}(t)) .
\end{equation}
 We require that $F(k)$ is a non-decreasing function of $k$.  
Notice that, in general, $\pc_{\I}(t)$ is a function not only of $k_{\I}(t)$ but also of $k_{j}(t)$ for all $j < \t$ because of the normalization, $Z$.  The attachment probabilities depend on all the node degrees unless $Z(t)$ is a function of $t$ alone.  This simpler form holds if and only if $F$ is a linear function, $F(k)=a+bk$. In the latter case, $Z(t)=(a+2b)t$ since $\sum_{j=0}^{t-1} k_{j}(t)=2t$.	

	The linear homogeneous case, $F(k)=k$ corresponds to the original Barabasi-Albert model~\cite{BaAl99} and leads to a scale free network where the degree distribution, $P(k)$, has a power law tail, $P(k) \sim k^{-3}$. More generally, if $F(k)$ is asymptotically linear, $P(k) \sim k^{\nu}$  where $\nu$ is tunable  to any value $2 \leq \nu < \infty$~\cite{DoMeSa00, KrReLe00, KrRe00}.   The asymptotically linear attachment kernel is a marginal case and marks a ``phase transition'' between regimes with qualitatively different behavior.  Consider the homogeneous models, $F(k)=k^{\al}$ studied in detail in Ref.\ \cite{KrRe00}.  In the sublinear case, $0<\al<1$ the degree distribution has a stretched exponential form and the node with the maximum degree has polylogarithmically many connections.  The limiting case of $\al=0$ is a random network where each connection is randomly and independently chosen.  There is an analogy between $\al$ and temperature in a thermodynamic system with the range  $0\leq \al <1$ like a high temperature phase. The order parameter is the maximum degree in the system divided by $N$ and the order parameter vanishes for $0 \leq \al \leq 1$.  In the superlinear or low temperature phase, $\al > 1$ there is a single, ``gel'' node that has almost all connections and the order parameter is unity.  The phase transition then has a discontinuous character despite the fact that the $\al=1$ state is scale free.  
	Another indication that the transition is discontinuous is seen by looking at the entropy.  Using the Kolmogorov-Chaitin definition of entropy as the minimum number of bits required to describe a system state~\cite{Mac99b}, it is clearly seen that the entropy per node is positive for all $\al \leq 1$ but that for $\al>1$ the entropy per node vanishes since almost all nodes connect to the gel node and it is only necessary to specify the connections for those nodes that do not connect to the gel node.  Thus, the entropy per node is also discontinuous at $\al=1$.

\section{Parallel computation and depth} 
\label{sec:comp}	
	Computational complexity theory is concerned with the scaling of computational resources needed to solve problems as a function of the size of the problem.  An introduction to the
field can be found in Ref.\ [\onlinecite{Papa}]. 
 Here we focus on parallel
computation and choose the standard {\em parallel random access machine} (PRAM) as
the model of computation~\cite{GiRy}.  The main resources of interest are {\em parallel time} or
{\em depth} and number of processors.  A PRAM consists of a number of simple
processors (random access machines or RAMs) all connected to a global memory. 
Although a RAM is typically defined with much less computational power than a real
microprocessor such as Pentium, it would not change the scaling found here to think of a
PRAM as being composed of many microprocessors all connected to the same random
access memory.   The processors run synchronously and each processor runs the same
program. Processors have an integer label so that different processors  follow different
computational paths.  
	The PRAM is the most powerful model of classical, digital computation.  The number of processors and memory is allowed to increase {\em polynomially} (i.e.\ as an arbitrary power) in the size of the problem to be solved.  Communication is non-local in that it is assumed that any processor can communicate with any memory cell in a single time step.  Obviously, this assumption runs up against speed of light or hardware density limitations.  Nonetheless, parallel time on a PRAM quantifies a fundamental aspect of computation.   Any problem that can be solved by a PRAM with $H$ processors in parallel time $T$ could also be solved by a single processor machine in a time $W$ such that $W \leq HT$ since the single processor could sequentially run through the tasks that were originally assigned to the $H$ processors.  The single processor time, $W$  is sometimes referred to as the computational work.  On the other hand, it is not obvious whether the work of a single processor can be re-organized so that it can be accomplished in a substantially smaller number of steps by many processors working independently during each step.  
		
An example of where exponential speed-up can be achieved through parallelism is adding $N$ numbers.  Addition can be done by a single processor in a time that scales linearly in $N$.  On a PRAM with $N/2$ processors addition can be carried out in ${\cal O}(\log
N)$ parallel time using a binary tree. For simplicity, suppose $N$ is a power of $2$.  In the first step, processor one adds the first and second numbers and puts the result in memory, processor two adds the third and fourth numbers and puts the result in memory and so on.  After the first step is concluded there are $N/2$ numbers to add and these are again summed in a pairwise fashion by $N/4$ processors.  The summation is completed after ${\cal O}(\log N)$ steps.  Addition is  said to have an {\em efficient} parallel algorithms in the sense that they can be solved in time that is a power of the logarithm of the problem size, here $N$, that is,  {\em polylog} time.  On the other hand, it is believed that there are some problems that can be solved in polynomial time using a single processor but cannot be efficiently parallelized.  It is believed that P-complete problems~\cite{GiRy,GrHoRu} have this property and cannot be solved in polylog time with polynomially many processors. 
	
The main concern of this paper is the complexity of generating networks defined by preferential attachment growth rules. Since these networks grow via a stochastic process, we envision a PRAM model equipped with registers containing random numbers.  The essential question that we seek to answer is the depth (number of PRAM steps) required to convert a set of independent random bits into a statistically correct network.

\section{Parallel Algorithms for Growing Network Models}
\label{sec:alg}

         At first glance, it seems that growing networks have a strong history dependence.  It would appear that to connect some node $t$ appropriately one must first connect all nodes prior to $t$ in order to compute the connection probabilities for $t$ according to Eq. \ref{eq:pat}. Surprisingly, one can construct a statistically correct network using an iterative parallel process that converges in far fewer than $t$ steps.  The strategy is to place progressively tighter lower bounds on the connection probabilities based on connections made in previous parallel steps in the process.
         
\subsection{A Coin Toss with Memory}

A simple example of the general strategy is instructive.  Consider a  biased coin toss with memory such that the number of heads on the first $t$ coin tosses modifies the probability of a head on toss $t+1$.  Suppose that more heads on previous tosses increases the probability of a head on the current toss according to some function $f(x)$ where $\pi(t)=f(x(t))$ is the probability of a head on the $t^{\rm th}$ coin toss and $x(t)$ is the fraction of heads on all the coins tossed before $t$.  Suppose that $f$ is a non-decreasing function of its argument and that $f(0)>0$.  Note that the special case $f(x)=x$ is a Polya Urn problem and is discussed in Sec.\ \ref{sec:redirect}.

The goal is to simulate a sequence of $N$ coin tosses. It would appear that we cannot decide coin $t$ until we have decided all its predecessors. Nonetheless, we can proceed in parallel by successively improving lower bounds on the probability that a given coin toss is a head.  Let, $p^S(t)$, 
\begin{equation}
p^S(t)=f(x^S(t))
\end{equation}
be an estimated lower bound on the probability that the $t^{\rm th}$ coin toss is a head on the $S^{\rm th}$ step of the algorithm where $x^S(t)$ is the fraction of tosses determined to be heads at the beginning of iteration $S$.  The starting assumption is that none of the tosses have been determined, $x^1(t)=0$ for all $t$, and this assumption is used to compute how many coins become heads on the first iteration.  Thus, $p^1(t)=f(0)$ and, on the first iteration, coin $t$ becomes a head with this probability.   Once a coin becomes a head, it stays a head while coins that are not heads remain undecided.  On the second iteration, we make use of the heads decided in the first iteration to recompute the fraction determined to be heads, $x^2(t)$ and from these obtain the new bounds  $p^2(t)=f(x^2(t)) \geq p^1(t)$.  For each coin $t$ that is not yet determined to be a head we declare it a head with conditional probability $\rho^2(t)$ that it will become a head on this step given that it is not yet a head,  $\rho^2(t)=(p^2(t)-p^1(t))/(1-p^1(t))$.  Some new coins are declared heads and these are then used to compute $x^3(t)$.  In general, if coin $t$ is not yet determined by step $S$, it becomes a head with probability 
\begin{equation}
\rho^S(t)=\frac{p^{S}(t)-p^{S-1}(t)}{1-p^{S-1}(t)} .
\end{equation}  
where $\rho^S(t)$ is the conditional probability of coin $t$ becoming a head on step $S$ given it was undecided up to step $S$.  The expression for the conditional probability follows from the observation that the denominator is the marginal probability of being undecided after step $S-1$ and the numerator is the probability of becoming a head on step $S$.
The algorithm stops on step $T$ when there is no change from one step to the next, $x^{T}(t)=x^{T-1}(t)$ for all  $t$, and the lower bounds equal the true probabilities $p^T(t)=\pi(t)$.  At the end of the simulation, every coin that is not a head is declared to be a tail.  For every $t$,
\begin{equation}
x^1(t) \leq x^2(t) \leq \ldots \leq  x^T(t)=x(t)
\end{equation}
so that
\begin{equation}
p^1(t) \leq p^2(t) \leq \ldots \leq  p^T(t)=\pi(t) .
\end{equation}
Thus the procedure is well-defined and we can decide in stages whether coin $t$ will be a head.  

In the following two sections we show how to generalize this strategy to the case of preferential attachment growing network models.  

\subsection{Parallel Algorithm for Linear and Sublinear Kernels}
\label{sec:subalg}
	This section describes a parallel algorithm for constructing a network with a sublinear or linear attachment rule, $F(k)=k^{\alpha}$ where $0 \leq \alpha \leq 1$ or, more generally, the case where the attachment weight $F(k)$ is a non-decreasing, convex function of $k$.  As in the coin toss example, on intermediate parallel steps we have nodes whose connections are not yet determined.  In this algorithm we lump all of these connections into a ``ghost'' node whose in-degree is equal to the number of nodes that have not yet been determined.  On every parallel time step, $\T$, the algorithm attempts to connect every node that is currently connected to the ghost node to a real node according to lower bounds on the connection probabilities determined by connections that have been made in previous steps.  

In the initialization, $\T=0$ step of the algorithm, a ghost node is created and all real nodes are connected to it, except node zero, which connects to itself, and node one, which also connects to  node zero. Thus, for $\T=0$ and every sequential time $t>1$, every real node $n<t$ has in-degree 0 and out-degree 1, except the zero node which has both in- and out-degree equal to 1.  The ghost node has in-degree $t-1$ for $t>0$.  Let  $k^{\T}_{g}(t)$ be the number of nodes connecting to the ghost node at the beginning of parallel step $\T$ and sequential time t so that  $k^{1}_{g}(t)=t-1$ for $t>0$.  In the first, $\T=1$ step of the algorithm the connection probability lower bound for node $t$ to connect to node $n$, $\w_n^1(t)$ is given by 
\begin{equation}
\w_n^1(t)=\left\{\begin{array}{cc}F(2)/\zt^1(t) & n=0 \\F(1)/\zt^1(t) & n>0\end{array}\right.
\end{equation}
while the connection probability, $\g^1(t)$ for the ghost node is taken to be proportional to its number of connections,
\begin{equation}
\label{eq:initghostprob}
\g^1(t)=c(t-1)/\zt^1(t)
\end{equation}
with the normalization given by
\begin{equation}
\label{eq:normdef}
\zt^1(t)=c(t-1) + F(2) + (t-1)F(1) .
\end{equation}
The constant $c$ is discussed below.  These are the connection probabilities that would arise if each real node has one connection and the ghost node has an attachment probability proportional to its degree.  On the first step of the algorithm, each node $t$ is connected to one of its predecessors or the ghost node according to the probabilities given above.

As in the case of the coin toss model described in the previous section, on successive steps we recompute the bounds on the connection probabilities $\w_n^\T(t)$ for the real nodes  and the ghost node $\g^\T(t)$.  For general $\T$, $t$ and $n$ these probabilities are given by
\begin{eqnarray}
\label{eq:wat}
 \w_n^\T(t) & = & F(k_n^\T(t))/\zt^\T(t)\\
 \label{eq:gwat}
 \g^\T(t)&=&ck_g^\T(t)/\zt^\T(t)
 \end{eqnarray}
with the normalization given by
\begin{equation}
\label{eq:ztilde}
\zt^\T(t)=ck_g^\T(t) + \sum_{m=0}^{t-1}F(k_m^\T(t)).
\end{equation}

On step $\T$ of the algorithm, the conditional probability, $\a_n^\T(t)$ of connecting node $t$ to node $n$, given that node $t$ has not yet connected to a real node on an earlier step, is given by the difference between the probability bounds on successive steps divided by the marginal probability of being undetermined (connected to the ghost node) before step $S$,
\begin{equation}
\a_n^\T(t)= \frac{ \w_n^\T(t)- \w_n^{\T-1}(t)}{\g^{\T-1}(t)} .
\end{equation}
Note that the denominator can be written as 
\begin{equation}
\g^{\T-1}(t)=1-\sum_{m=0}^{t-1}\w_{m}^{\T-1}(t).
\end{equation}
On step $\T$ of the algorithm each node $t$ that was still connected to the ghost node after step $\T-1$ is connected with probability $\a_n^\T(t)$ to real node $n<t$ or, with probability, $\a_g^\T(t)$,
\begin{equation}
\label{eq:ghostprob}
\a_g^\T(t)=\g^\T(t)/\g^{\T-1}(t)
\end{equation}
still connected to the ghost node.  The algorithm is finished after $T$ steps when there are no more nodes connected to the ghost node and the bounds of Eq.\ \ref{eq:wat} saturate to the correct probabilities of Eq.\  \ref{eq:pat}.  Note that at least one node must connect in each parallel step since the lowest numbered node that is still unconnected will have no weight allotted to it in the ghost node. 

For the conditional probabilities $\a_n^\T(t)$ to be positive, the  probability bounds must be non-decreasing for all $n$ and $t$,
\begin{equation}
\label{eqn:mono}
\w_n^1(t) \leq \w_n^2(t) \leq \cdots \leq \w_n^T(t) = \pc_n(t).
\end{equation}
These inequalities imply a bound on $c$ as follows. Since $F(k)$ is a non-decreasing function of $k$ and $k_n^\T(t)$ is a non-decreasing of $\T$ it is sufficient to require that $\zt(t)$ is a non-increasing function, $\zt^\T(t) \leq \zt^{\T-1}(t)$, or, since,
\begin{equation}
k_g^{\T}(t) = 2t-1- \sum_{m=0}^{t-1}k_m^{\T}(t)
\end{equation}
we require that
\begin{equation}
\sum_{m=0}^{t-1} (F(k_m^\T(t))-ck_m^\T(t)) \leq \sum_{m=0}^{t-1} (F(k_m^{\T-1}(t))-ck_m^{\T-1}(t)).
\end{equation}
This inequality is satisfied term by term if $F(k)-ck$ is non-increasing which holds if 
\begin{equation}
\label{eq:cbound }
c \geq \max_k \{F(k+1)-F(k)\} .
\end{equation}
Since the algorithm will finish fastest if the ghost node has the smallest possible weight, we set $c$ equal to its lower bound. In particular, for the power law case, $F(k)=k^\alpha$ with $\alpha \leq 1$, the maximum occurs for $k=1$ yielding
\begin{equation}
\label{eq:cvalpha}
c= 2^\alpha - 1
\end{equation}

\subsection{The Parallel Algorithm for Superlinear Kernels}

	For the superlinear case, a gel node develops to which almost all nodes  connect as $N \rightarrow  \infty$.  When $\alpha > 2$ all but a finite number of nodes connect to the gel node.  The parallel algorithm described here takes advantage of the fact that the vast majority of connections are to the gel node and the gel node plays a role similar to that of the ghost node in the sublinear and linear cases.  The basic structure of the algorithm is as follows.  In the initialization $\T=0$ phase the sequential algorithm is run so that all nodes $t \leq \s$ are properly connected. $\s$ is chosen so that a single gel node is firmly established by the time all nodes $t \leq \s$ are connected. The gel node is firmly established if the probability that a different node ultimately becomes the gel node is less than some small value $\epsilon$.  When $\alpha$ is large $\s$ is small, but as $\alpha$ approaches 1, for fixed $\epsilon$, $\s$ diverges.  After the initialization phase, it is tentatively assumed that all nodes $\s< t < N$ are connected to the gel node.  The gel node serves as a repository for all connections that are not yet determined. In successive steps, the connection probabilities of all nodes $\s< t < N$ are modified according to the number of connections that possible destination nodes, $n<t$ received in the previous step and lower bounds on connection probabilities are recalculated.  The difference between old and new probability bounds are used to find conditional probabilities for moving a connection from the gel node to some other node.  This process is repeated until no connections are moved from the gel node to any other node. The nodes that have not been moved away from the gel node are then determined to be connected to the gel node.

Following the general strategy, lower bounds on the connection probabilities for $t>\s$ are determined for each parallel step,
\begin{equation}
\label{eq:prob}
\w_{n}^{\T}(t)=\frac{F(k_{n}^{\T}(t))}{Z^\T(t)}
\end{equation}
where the normalization is given by 
\begin{equation}
\label{eqn:norm}
Z^\T(t)=\sum_{n=0}^{t-1}F(k_{n}^{\T}(t)).
\end{equation}
Note that the connection probabilities are calculated in the same way for the gel node and the other nodes in contrast to the sublinear case.   

In the first parallel step, $\T=1$, the algorithm connects every node, $t>\s$ to some node, $n$ according to the connection probabilities $\w_{n}^{1}(t)$.  In successive, steps, $\T > 0$, it attempts to re-connect only those nodes $t>\s$ that are still connected to the gel node.  The conditional probability $\a_n^S(t)$ for connecting $t>\s$ to $n \neq g$ on step $S$ is given by
\begin{equation}
\label{eq:conprob}
\a_{n}^{\T}(t)=\frac{\w_{n}^{\T}(t)-\w_{n}^{\T-1}(t)}{\w_{g}^{\T-1}(t)}  .
\end{equation}
The numerator is the probability that $t$ connects to $n$ on step $\T$ and the denominator is the probability that $t$ is undetermined after step $\T-1$. The conditional probability that $t$ is undetermined after step $\T$, given that it was undetermined after step $\T-1$, is
\begin{equation}
\a_{g}^{\T}(t)=\frac{\w_{g}^{\T}(t)}{\w_{g}^{\T-1}(t)}. \label{eq:sizeofgel}
\end{equation}
The algorithm is finished after step $T$ if  no changes occur from step $T-1$ to step $T$.  On step $T$ nodes that are connected to $g$ are considered to be determined and actually connected to the gel node.

The algorithm is valid if Eq.\ \ref{eqn:mono} holds for all $t>\s$ and $n \neq g$.  Since $F(k)$ is non-decreasing, we require that $Z^\T(t)$ is a non-increasing function of $\T$.  From Eq.\  \ref{eqn:norm} we must show that the change in $Z$ from one parallel step to the next is either constant or decreasing for all $t$ and $\T$.  We can write the requirement for the validity of the algorithm as
\begin{equation}
\label{eqn:Zdecrease}
Z^{\T}-Z^{\T-1}=\sum_{m=0}^{t-1}[F(k_m^{\T})-F(k_m^{\T-1})] \leq 0.
\end{equation}
It is useful to take the gel node term out of the sum, as its behavior is different 
\begin{equation}
\label{eqn:Zdecrease2}
Z^{\T}-Z^{\T-1}=F(k_g^{\T})-F(k_g^{\T-1})+\sum_{m=0,m \neq g}^{t-1}[F(k_m^{\T})-F(k_m^{\T-1})].
\end{equation}
At each parallel step connections are switched from the gel node to other nodes.  For every connection that is lost by the gel node exactly one connection is gained by another node. We also note that because $F$ is a concave function with a continuously increasing derivative, we can say that for any positive $\delta k$
\begin{equation}
\label{eqn:deltak}
F^{\prime}(k+\delta k) \delta k \geq F(k+\delta k)-F(k) \geq F^{\prime}(k) \delta k
\end{equation}
Since $k_g^{\T}$ is decreasing with $\T$ and $k_{m \neq g}^{\T}$ is increasing with $\T$ we can rewrite Eq.\ \ref{eqn:Zdecrease2}, absorbing the contribution of the gel node into the sum to describe the entire rewiring of $k_m^{\T}-k_m^{\T-1}$ connections from the gel node to another node.  We use eq.\ \ref{eqn:deltak} to put an upper bound on the change in size, the RHS of Eq.\ \ref{eqn:Zdecrease2}

\begin{equation}
\label{eqn:Zdecrease3}
Z^{\T}-Z^{\T-1} \leq \sum_{m=0,m \neq g}^{t-1}[k_m^{\T}-k_m^{\T-1}][F^{\prime}(k_m^{\T})-F^{\prime}(k_g^{\T})].
\end{equation}
The term on the RHS in the first square brackets is non-negative.  If $k_g^{\T} > k_{m}^{\T}$ then the term in square brackets is always negative because $F^{\prime}$ is a strictly increasing function of $k$.  This argument shows that Eq.\ \ref{eqn:Zdecrease} holds and thus the algorithm is valid if the gel node remains the largest node until the end of the simulation.  The value of $\epsilon$ and, thus, the choice of $\s$ determines the error rate of the algorithm since the algorithm fails if and only if the gel node loses it status as having the most connections.

\subsection{Redirection Method for Linear Kernels and Urns}
\label{sec:redirect}

	This section explores the method proposed by Krapivsky and Redner\cite{KrRe00} for the case of a linear attachment kernel.  We show that this method can be used to generate the network in $\ord (\log \log N)$ steps.  The method works as follows:  At sequential time $t$, node $t$ is connected to any node $\I<t$ with equal probability.  With probability $r$, however, this node is redirected to the ``ancestor'' of $\I$, the node that $\I$ connects to.  As Krapivsky and Redner show, when $r=0.5$, this procedure exactly reproduces the \AB model ($F(k)=k$). For other values of $r$, $F(k)$ is asymptotically linear and the connectivity distribution scales as $P_{k} \sim k^{-\nu}$ where $\nu = 1+1/r$.  It is easy to see why redirection is equivalent to a linear kernel.  A node that already has $k$ connections has $k$ ways to be connected from a new node since each of the $k$ connections can serve as a redirection point for the new node.  For  $r=(1-r)=1/2$ it is clear that $F(k)=k F(1)$ so this case corresponds to homogeneous \AB network.
	
This redirection process can be simulated in $\ord(\log \log N)$ parallel time as follows.  First, randomly connect every node to one of its predecessors.  Once this is done, for every connection, with probability $r$, make that connection a redirectable (R) connection, otherwise, make it a terminal (T) connection.  All that remains is to trace every path of R connections until a T connection is reached.  This can be accomplished using a standard parallel connectivity algorithm or by the following simple approach.  For every node, $t$, if its outgoing connection is type T make no change to the connection.  If its outgoing connection is type R, then it is redirected.  Suppose $t$ connects to $i$ by an R connection and that $i$ connects to $j$, then after the parallel step, $t$ connects to $j$.  Furthermore, if the $i$ to $j$ connection is type T then the new connection from $t$ to $j$ is type T, otherwise it is an R connection.  When all of the connections are type T, the program is done and the network is correctly wired. It is clear that this procedure require a number of steps that scales as the logarithm of the longest chain of redirections.  On average, the longest chain of redirections will behave as the logarithm of the system size.  Each connection redirects with probability $r$.  The average length of the longest chain of redirections, $M$, is estimated by $Nr^{M} \approx 1$ where $N$ is the number of possible starting points and $r^M$ is the probability of a chain of length $M$.  Thus $\log N+M \log r \approx 0$ so $M \sim -\log N/\log r$. Note that the chain length saturates at $\ord(\log N)$ rather than diverges as $r \rightarrow 1$.  Even if $r \rightarrow 1$ each connection will typically halve the distance to the origin so that  there are $\ord (\log N)$ connections in the longest chain. A chain of connections of length $M$, can be traced in $\log M$ steps, because each step will halve the length of the chain.  Thus the algorithm will finish in ${\cal O}( \log M)=\ord (\log \log N)$ steps.  

The Polya urn model is closely related to the \AB growing network model and the redirection method can be applied to efficiently sample  its histories in parallel.  In the simplest version of the model, an urn initially contains two balls, one red and one black.  On each time step a ball is randomly selected from the urn and then replaced along with a new ball of the same  color.  Thus, after $N$ steps the urn contains $N+2$ balls. The urn model has the unusual property that it can have any limit law.  For large $N$ the fraction of red balls approaches a constant but, with equal probability, that constant can be any value from zero to one.  The limit law is thus determined by the initial choices.   

The urn model can be viewed as a network where each ball is a node and the connection from one node to a predecessor represents the fact that the color of the later node was determined by the earlier node.  To find the color of a given ball or node, the connections are traced back to one of the two initial balls.  This representation shows that the urn model is identical to the linear network model in the limit that the redirection probability is unity.  The typical longest path of connections back to the origin is $\ord (\log N)$ since each connection will typically halve the distance to the origin. Thus the depth of sampling the history of an urn model is $\ord (\log \log N)$.

\section{Efficiency of Parallel Algorithms}
\label{sec:eff}
\subsection{Efficiency of the Parallel Algorithm when $0 \leq \alpha \leq 1$}
\label{sec:effsub}
	In this section we argue that for a system of size $N$, when $0 \leq \alpha \leq 1$, the parallel algorithm will finish in ${\cal O} (\log N)$ parallel steps and we estimate the prefactor of the logarithm.
The starting point is an equation for the expected number of connections to the ghost node on the $\T+1$ step given the number of connections on steps $\T$ and $\T-1$,  
\begin{equation}
\label{eq:kgvs}
{\bf E}(k_{g}^{\T+1}(t))=\sum_{t^{ \prime}=1}^{t-1}[k_{g}^{\T}(t^{\prime})-k_{g}^{\T}(t^{\prime}-1)]\a_g^{\T}(t^{\prime}).
\end{equation}
The quantity in the square brackets is 1 if $t^\prime$ is connected to the ghost node on step $\T$ and 0 otherwise while $\a_g^{\T}(t^{\prime})$, defined in Eq.\ \ref{eq:ghostprob} is the conditional probability that $t^\prime$ connected to the ghost node after step $\T$ if it is connected to the ghost node before step $\T$.   Equation \ref{eq:kgvs} holds for $\T>2$.  Initially, $k_g^1(t)=t-1$. Specializing to the case that the attachment kernel is a pure power law with exponent $\alpha$ and ignoring constants that are irrelevant in the large $t$ limit we have  
\begin{equation}
\label{eq:kginit}
{\bf E}(k_g^2(t))= c/(c+1)= 1-2^{-\alpha}.
\end{equation}
This result follows from the fact that the probability that node $t$ will still be connected to the ghost node after the first step is, according to Eqs.\ \ref{eq:initghostprob} and \ref{eq:normdef}, approximately $c/(c + 1)$.  The far RHS of the expression is obtained from Eq.\ \ref{eq:cvalpha}.

To proceed further we make two approximations.  First, we ignore fluctuations and replace $k_g$ by its average value on the RHS of Eq.\ \ref{eq:kgvs},
\begin{equation}
\label{eq:kgvs1}
k_{g}^{\T+1}(t)=\sum_{t^{ \prime}=1}^{t-1}[k_{g}^{\T}(t^{\prime})-k_{g}^{\T}(t^{\prime}-1)] \frac{k_g^{\T}(t^{\prime})} {k_g^{\T-1}(t^{\prime}) }\frac{ \zt^{\T-1}(t^{\prime})}{\zt^{\T}(t^{\prime})}
\end{equation}
where the notation is simplified in this equation by interpreting $k_g$ as the average number of connections to the ghost node and where Eqs.\ \ref{eq:gwat} and \ref{eq:ghostprob} have been used to expand $\a_g$.

For the case of a linear attachment kernel, $c=1$ and the normalization $\zt^{\T}$ is independent of $\T$. The ratio of normalizations thus drops out of the equation and we obtain,
\begin{equation}
\label{eq:kgvs2}
k_{g}^{\T+1}(t)=\sum_{t^{ \prime}=1}^{t-1}[k_{g}^{\T}(t^{\prime})-k_{g}^{\T}(t^{\prime}-1)] \frac{k_g^{\T}(t^{\prime})} {k_g^{\T-1}(t^{\prime}) }
\end{equation}
For sublinear kernels, the choice of $c$ insures that the ratio  $\zt^{\T-1}(t^{\prime})/\zt^{\T}(t^{\prime})$ is less than one as discussed at the end of Sec.\ \ref{sec:subalg}. Our second approximation, is to assume that this ratio is unity for the entire sublinear regime. 
Note that both $k_g^1(t)$ and $k_g^2(t)$ are proportional to $t$.  It follows from Eq.\ \ref{eq:kgvs2} that $k_g^\T(t)$ is is proportional to  $t$ for all $\T$ and we write $k_g^\T(t)=\kappa(\T)t$.  This substitution reduces Eq.\ \ref{eq:kgvs2} to 
\begin{equation}
\label{eq:kgvs3}
\kappa(\T+1)= \frac{\kappa(\T)^2}{\kappa(\T-1)}
\end{equation}
Given our approximations, the ratio $\kappa(\T)/\kappa(\T-1)=1-2^{-\alpha}$ for all $\T$ and the solution is $\kappa(\T)=(1-2^{-\alpha})^\T$.  The estimate for the number of steps, $T$ needed to complete the algorithm is such that the ghost node is expected to have fewer than one node, $\kappa(T)N=1$.  This requirement leads to the result
\begin{equation}
\label{eq:subtheory}
T= \frac{\log(N)}{-\log(1-2^{-\alpha})} .
\end{equation}
This result is compared to the numerical simulations in Sec.\ \ref{sec:sim}.

 \subsection{Efficiency of the parallel algorithm when $\alpha > 1$}
 
 In this section we show that the  $\alpha >1$ algorithm finishes in constant time independent of $N$ although this constant diverges as $\alpha \rightarrow 1$.  The key fact \cite{KrRe00} about superlinear networks is that there is a cut-off 
\begin{equation} 
\label{eq:kmax}
k_{\rm max}= \alpha/(\alpha -1)
\end{equation} 
such that only a finite number of nodes have more than $k_{\rm max}$ connections.  By choosing $\s$ sufficiently large, no nodes $t > \s$ will have more than $k_{\rm max}$ connections.  We will show that the running time of the parallel part of the algorithm is roughly $k_{\rm max}$ steps.  
  
Consider what happens on the first step of the algorithm.  All nodes $t>\s$ are initially connected to the gel node so the leading behavior of the normalization is  $ Z^1(t) \sim t^\alpha$ and the leading behavior of the connection probabilities, defined in Eq.\ \ref{eq:prob}, is 
\begin{equation}
\w_n^1(t) \sim t^{-\alpha}   
\end{equation}
for $n \neq g$. Summing over all $n \neq g$ we find that on the first step, the probability that node $t$ will connect away from the gel node behaves as $t^{1-\alpha}$.  The expected change in the number of nodes connecting to the gel node during step one is obtained by summing over all nodes $\s < t < N$, with the result
\begin{equation}
\label{eq:deltak2}
\delta k_g^2(N) \equiv k_g^1(N)-k_g^2(N) \sim N^{2-\alpha}.
\end{equation}
If $\alpha>2$ no changes are expected to occur in the first step of the algorithm and we are done.  This result is consistent with the fact  that for $\alpha>2$, there are only a finite number of nodes with more than one connection and these are all determined before $\s$. 

For $1 < \al < 2$ additional steps are needed before $\delta k_g^\T(N)$ is less than one and the algorithm is done.  We make the ansatz that
\begin{equation}
\label{eq:ansatz}
\delta k_g^{\T}(t) \sim t^{\gamma(\T)}
\end{equation}
and look for a self-consistent solution for $\gamma(\T)$.  The running time, $T$  is obtained by solving for the least $T$ such that $\gamma(T)<0$.

On the second and later steps of the algorithm, the conditional connection probabilities, defined in Eqs.\ \ref{eq:prob} to \ref{eq:conprob}, can be written to leading order and for $n \neq g$ as,
\begin{equation}
\a_n^\T(t) =\frac{(Z^{\T-1}(t)/Z^{\T}(t)) k_{n}^{\T}(t)^\al -k_{n}^{\T-1}(t)^\al}{k_{g}^{\T-1}(t)^\alpha}
\end{equation}
There are two ways for $\a_n^\T(t)$ to be non-zero.  The first is for there to have been a new connection from $t^\prime$ to $n$, with $n< t^\prime<t$,  in step $\T-1$.  The expected number of nodes, $n$ that received new connections in step $\T-1$ is just $\delta k_g^\T(t)$.  Since $k_g^\T(t) \sim t$ for all $\T$, the leading behavior of $\a_n^\T(t)$ is $t^{-\alpha}$ and the overall probability that $t$ will connect away from the gel node by this mechanism  scales as $\delta k_g^\T(t) t^{-\alpha} \sim t^{\gamma(\T)-\al}$. 

The second way for $\a_n^\T(t)$ to be non-zero is for the ratio $Z^{\T-1}(t)/Z^{\T}(t)$ to exceed unity.  This possibility applies for all target nodes, $n<t$, $n \neq g$. The leading behavior of this ratio is given by
\begin{equation}
Z^{\T-1}(t)/Z^{\T}(t) \sim \frac{(k_g^\T(t) + \delta k_g^\T(t))^\alpha}{k_g^\T(t)^\alpha} \sim 1+ \alpha \frac{\delta k_g^\T(t)}{k_g^\T(t)}
\end{equation}
so that the leading behavior of $\a_n^\T(t)$ is $\delta k_g^\T(t) t^{-(1+\alpha)}$.  Since there are $t$ target nodes, the total probability that $t$ will connect away from the gel node by this mechanism  again scales as $ t^{\gamma(\T)-\alpha}$. 

Combining both of the above mechanisms for $t$ to connect away from the gel node and summing over all $t$, $\s< t< N$ (and still connected to the gel node) we obtain an expression for $\delta k_g^{\T+1}(N)$, the expected number of nodes directed away from the gel node on step $\T$,
\begin{equation}
\delta k_g^{\T+1}(N) \sim N^{\gamma(\T)+ 1-\alpha}
\end{equation}
Using the ansatz of Eq.\ \ref{eq:ansatz} we obtain the recursion relation,
\begin{equation}
\gamma(\T+1) =  \gamma(\T) + 1 - \al .
\end{equation}
The recursion relation and the initial condition $\gamma(2)=2-\al$, Eq.\ \ref{eq:deltak2},  has the solution, 
\begin{equation}
\gamma(\T)=\al - (\al-1)\T .
\end{equation}
The running time of the algorithm is obtained from the least $T$ for which $\gamma(T)$ is negative,
\begin{equation}
 T = \al/(\al-1)
 \end{equation}
 or, from, Eq.\ \ref{eq:kmax},
\begin{equation}
T  = k_{\rm max} .
\end{equation}
This result can be understood in terms of the following sequence of events for creating connections for nodes beyond $\s$.  In the first parallel step almost all nodes with two connections are generated. In the second parallel step a small fraction of these nodes develop a third connection and a comparable number of nodes with one connection get a second connection.  On the third step, an even smaller number of nodes with three connections get a fourth connection and so on until nothing happens.  Note that the analysis of the algorithm reproduces the results in \cite{KrRe00} for the scaling of the number of nodes with $2, 3, \ldots k_{\rm max}$ connections.

\section{Simulation Results for Linear and Sublinear Kernels}
\label{sec:sim}

\begin{figure}
\includegraphics[width=3in]{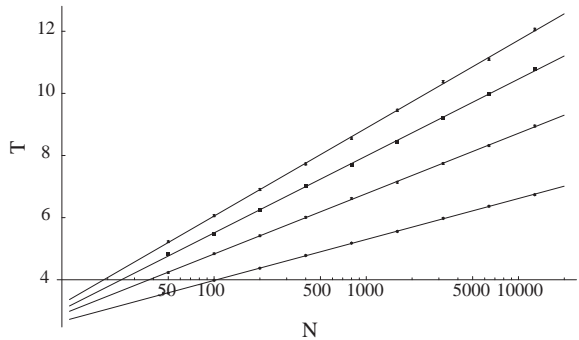}\caption{The average parallel time $T$ to generate a network as a function of system size $N$ for $\alpha=0.25$, $0.5$, $0.75$ and $1.0$, from bottom to top, respectively.}
\label{fig:timevsize}
\end{figure}

\begin{figure}
\includegraphics[width=3in]{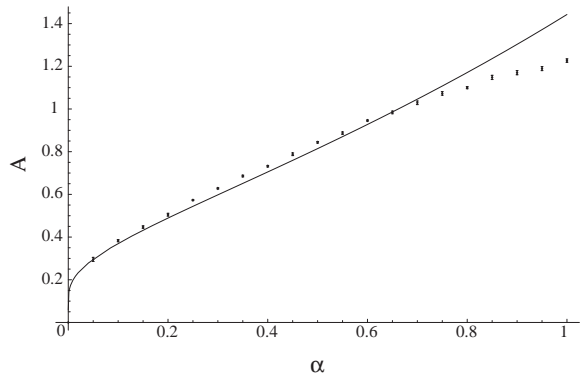}\caption{The coefficient $A$ of the leading logarithmic term in the running time versus $\alpha$.  The points are the results of the simulation and the solid line is the theoretical approximation, Eq.\ \ref{eq:subtheory}.}
\label{fig:coeffvalpha}
\end{figure}

	In Sec.\ \ref{sec:effsub} we argued that the algorithm for the sublinear kernel requires logarithmic parallel time to generate a network and in Eq.\ \ref{eq:subtheory} we  estimate the coefficient of the logarithm.  In this section we support these conclusions with a simulation of the parallel algorithm on a single processor workstation.  In the simulation the work of each processor on the PRAM is done in sequence making sure not to update the database describing the network until a parallel step is completed.
	We generated 1000 networks for each value of $\al$ and for each system size.  Values of alpha ranged from 0 to 1, in increments of 0.05 and system sizes from 50 nodes to 12,800 nodes with each size separated by a factor of two.   Figure \ref{fig:timevsize} shows the average number of parallel steps vs.\ system size  for $\alpha=$ 0.25, 0.5, 0.75 and 1.0.  The figure demonstrates the logarithmic dependance of average running time, $T$ on system size, $N$ for all values of $\al$ and the full range of system sizes so that that, to good approximation, $T = A(\alpha)\log N$.
	Figure \ref{fig:coeffvalpha} shows a plot of the coefficient $A$ as a function of  $\alpha$.  The results are plotted for $0 \leq \alpha \leq 1$.  The prediction of Eq.\ \ref{eq:subtheory} is shown on the same figure.  Although not perfect, the approximation of Eq.\ \ref{eq:subtheory} captures the general trend of the data and is within a few percent of the numerical results for $\al<0.8$.  The larger fluctuations in connectivity near $\al=1$ may explain why the ``mean field'' assumption underlying the theoretical curve loses accuracy there.  The theoretical estimate does appear to correctly predict that $A(\al)$ approaches zero with infinite slope as $\al \rightarrow 0$.

\section{Discussion}
\label{sec:con}	
	We have examined the parallel computational complexity of  generating networks obeying preferential attachment growth rules.  We demonstrated that these networks can be sampled in parallel time that is much less than the size of the network.  This result is surprising because the defining rules for generating these networks are sequential with nodes added to the network one at a time depending on the present state of the network.  Nonetheless, we have bounded the depth of sampling growing networks by exhibiting efficient parallel algorithms for the three cases, $0 \leq \al < 1$, $\al=1$ and $\al >1$.  The average parallel running time for the $0 \leq \al < 1$ algorithm is logarithmic, the algorithm for the \AB scale free network runs in $\ord(\log \log N)$ time and for $\al >1$ the algorithm runs in constant time. 
	 
	Growing networks thus provide an example of a discontinuous phase transition in complexity as a function of  $\al$ at $\al=1$.  It is not surprising that a complexity transition occurs at $\al=1$ since this is where the structural properties of the system also undergo a discontinuous transition from a high temperature ($\al<1$) regime where no nodes have a finite fraction of the connections to a low temperature ($\al>1$) regime where there is a single gel node with almost all connections.  It is noteworthy that parallel time is the proper resource to observe this transition.  The more common complexity measure of sequential time or computational work has no transition since it requires $\ord(N)$ time to give an explicit description of the network for any $\al$.
	
	Our results set upper bounds on the depth of sampling growing networks but we cannot rule out the existence of yet faster parallel algorithms.  For example, if a constant time algorithm exists for $0 \leq \al < 1$, it would modify the conclusion that there is a discontinuous complexity transition at $\al=1$.  There are few rigorous lower bounds in computational complexity theory, so, in general, conclusions concerning the depth of sampling  and the existence of complexity transitions in statistical physics must be considered tentative.
	
In this paper we have presented a general strategy for parallelizing a broad class of sequential stochastic processes, exemplified by the coin toss with memory.  We have applied the general method to create algorithms that efficiently parallelize  preferential attachment network models.  The general method should be more broadly applicable to growing network models with more complicated rules.  To give one example, Hajra and Sen~\cite{HaSe04} extend the preferential attachment model to include an aging factor ($F(k)$ becomes $F(k,t-n)$) so that older nodes are either favored or avoided depending on a parameter.  Our algorithm can be modified to efficiently handle this class of models.  

It is also instructive to examine a growing network model where our general method is not efficient.  If $\al<0$, a case examined by Onody and deCastro~\cite{OnCa04}, the general method can be applied but will not be efficient. The problem is that lower bounds on connection probabilities are typically extremely small and the algorithm will connect only a few nodes in each parallel step.  We are currently investigating methods to efficiently parallelize $\al<0$ networks.  

The fact that preferential attachment growing networks have no more than logarithmic depth indicates that they are not particularly complex objects.  On the other hand, very complex biological and social systems generate networks with similar properties.  If growing network models  accurately describe the networks generated by these systems one most conclude that the complexity and history dependence of the systems generating the networks are not manifest in the networks themselves.  An alternative possibility is that the real networks are themselves complex but that growing network models lack some essential statistical properties of the real networks.

\acknowledgements
This work was supported by NSF grant DMR-0242402.


\begin{thebibliography}{17}
\expandafter\ifx\csname natexlab\endcsname\relax\def\natexlab#1{#1}\fi
\expandafter\ifx\csname bibnamefont\endcsname\relax
  \def\bibnamefont#1{#1}\fi
\expandafter\ifx\csname bibfnamefont\endcsname\relax
  \def\bibfnamefont#1{#1}\fi
\expandafter\ifx\csname citenamefont\endcsname\relax
  \def\citenamefont#1{#1}\fi
\expandafter\ifx\csname url\endcsname\relax
  \def\url#1{\texttt{#1}}\fi
\expandafter\ifx\csname urlprefix\endcsname\relax\def\urlprefix{URL }\fi
\providecommand{\bibinfo}[2]{#2}
\providecommand{\eprint}[2][]{\url{#2}}

\bibitem[{\citenamefont{Newman}(2001)}]{Ne00}
\bibinfo{author}{\bibfnamefont{M.}~\bibnamefont{Newman}},
  \bibinfo{journal}{Proc. Natl. Acad. Science. USA}
  \textbf{\bibinfo{volume}{98}}, \bibinfo{pages}{404} (\bibinfo{year}{2001}).

\bibitem[{\citenamefont{Albert and Barabasi}(1999)}]{BaAl99}
\bibinfo{author}{\bibfnamefont{R.}~\bibnamefont{Albert}} \bibnamefont{and}
  \bibinfo{author}{\bibfnamefont{A.-L.} \bibnamefont{Barabasi}},
  \bibinfo{journal}{Science} \textbf{\bibinfo{volume}{286}},
  \bibinfo{pages}{509} (\bibinfo{year}{1999}).

\bibitem[{\citenamefont{Krapivsky et~al.}(2000)\citenamefont{Krapivsky, Redner,
  and Leyvraz}}]{KrReLe00}
\bibinfo{author}{\bibfnamefont{P.~L.} \bibnamefont{Krapivsky}},
  \bibinfo{author}{\bibfnamefont{S.}~\bibnamefont{Redner}}, \bibnamefont{and}
  \bibinfo{author}{\bibfnamefont{F.}~\bibnamefont{Leyvraz}},
  \bibinfo{journal}{Phys. Rev. Lett.} \textbf{\bibinfo{volume}{85}},
  \bibinfo{pages}{4629} (\bibinfo{year}{2000}).

\bibitem[{\citenamefont{Krapivsky and Redner}(2001)}]{KrRe00}
\bibinfo{author}{\bibfnamefont{P.}~\bibnamefont{Krapivsky}} \bibnamefont{and}
  \bibinfo{author}{\bibfnamefont{S.}~\bibnamefont{Redner}},
  \bibinfo{journal}{Phys. Rev. E} \textbf{\bibinfo{volume}{63}},
  \bibinfo{pages}{066123} (\bibinfo{year}{2001}).

\bibitem[{\citenamefont{Machta and Greenlaw}(1994)}]{MaGr}
\bibinfo{author}{\bibfnamefont{J.}~\bibnamefont{Machta}} \bibnamefont{and}
  \bibinfo{author}{\bibfnamefont{R.}~\bibnamefont{Greenlaw}},
  \bibinfo{journal}{J. Stat. Phys.} \textbf{\bibinfo{volume}{77}},
  \bibinfo{pages}{755} (\bibinfo{year}{1994}).

\bibitem[{\citenamefont{Machta and Li}(2001)}]{MaLi01}
\bibinfo{author}{\bibfnamefont{J.}~\bibnamefont{Machta}} \bibnamefont{and}
  \bibinfo{author}{\bibfnamefont{X.-N.} \bibnamefont{Li}},
  \bibinfo{journal}{Physica A} \textbf{\bibinfo{volume}{300}},
  \bibinfo{pages}{245} (\bibinfo{year}{2001}).

\bibitem[{\citenamefont{Moore and Machta}(2000)}]{MoMa00}
\bibinfo{author}{\bibfnamefont{C.}~\bibnamefont{Moore}} \bibnamefont{and}
  \bibinfo{author}{\bibfnamefont{J.}~\bibnamefont{Machta}},
  \bibinfo{journal}{J. Stat. Phys.} \textbf{\bibinfo{volume}{99}},
  \bibinfo{pages}{661} (\bibinfo{year}{2000}).

\bibitem[{\citenamefont{Machta and Greenlaw}(1996)}]{MaGr96}
\bibinfo{author}{\bibfnamefont{J.}~\bibnamefont{Machta}} \bibnamefont{and}
  \bibinfo{author}{\bibfnamefont{R.}~\bibnamefont{Greenlaw}},
  \bibinfo{journal}{J. Stat. Phys.} \textbf{\bibinfo{volume}{82}},
  \bibinfo{pages}{1299} (\bibinfo{year}{1996}).

\bibitem[{\citenamefont{Tillberg and Machta}(2004)}]{TiMa04}
\bibinfo{author}{\bibfnamefont{D.}~\bibnamefont{Tillberg}} \bibnamefont{and}
  \bibinfo{author}{\bibfnamefont{J.}~\bibnamefont{Machta}},
  \bibinfo{journal}{Phys. Rev. E} \textbf{\bibinfo{volume}{69}},
  \bibinfo{pages}{051403} (\bibinfo{year}{2004}).

\bibitem[{\citenamefont{Monasson et~al.}(1999)\citenamefont{Monasson, Zecchina,
  Kirkpatrick, Selman, and Troyansky}}]{MoZeKiSeTr99}
\bibinfo{author}{\bibfnamefont{R.}~\bibnamefont{Monasson}},
  \bibinfo{author}{\bibfnamefont{R.}~\bibnamefont{Zecchina}},
  \bibinfo{author}{\bibfnamefont{S.}~\bibnamefont{Kirkpatrick}},
  \bibinfo{author}{\bibfnamefont{B.}~\bibnamefont{Selman}}, \bibnamefont{and}
  \bibinfo{author}{\bibfnamefont{L.}~\bibnamefont{Troyansky}},
  \bibinfo{journal}{Nature} \textbf{\bibinfo{volume}{400}},
  \bibinfo{pages}{133} (\bibinfo{year}{1999}).

\bibitem[{\citenamefont{Dorogovtsev et~al.}(2000)\citenamefont{Dorogovtsev,
  Mendes, and Samukhin}}]{DoMeSa00}
\bibinfo{author}{\bibfnamefont{S.~N.} \bibnamefont{Dorogovtsev}},
  \bibinfo{author}{\bibfnamefont{J.~F.~F.} \bibnamefont{Mendes}},
  \bibnamefont{and} \bibinfo{author}{\bibfnamefont{A.~N.}
  \bibnamefont{Samukhin}}, \bibinfo{journal}{Phys. Rev. Lett.}
  \textbf{\bibinfo{volume}{85}}, \bibinfo{pages}{4633} (\bibinfo{year}{2000}).

\bibitem[{\citenamefont{Machta}(1999)}]{Mac99b}
\bibinfo{author}{\bibfnamefont{J.}~\bibnamefont{Machta}}, \bibinfo{journal}{Am.
  J. Phys.} \textbf{\bibinfo{volume}{67}}, \bibinfo{pages}{1074}
  (\bibinfo{year}{1999}).

\bibitem[{\citenamefont{Papadimitriou}(1994)}]{Papa}
\bibinfo{author}{\bibfnamefont{C.~H.} \bibnamefont{Papadimitriou}},
  \emph{\bibinfo{title}{Computational Complexity}} (\bibinfo{publisher}{Addison
  Wesley}, \bibinfo{year}{1994}).

\bibitem[{\citenamefont{Gibbons and Rytter}(1988)}]{GiRy}
\bibinfo{author}{\bibfnamefont{A.}~\bibnamefont{Gibbons}} \bibnamefont{and}
  \bibinfo{author}{\bibfnamefont{W.}~\bibnamefont{Rytter}},
  \emph{\bibinfo{title}{Efficient Parallel Algorithms}}
  (\bibinfo{publisher}{Cambridge University Press}, \bibinfo{year}{1988}).

\bibitem[{\citenamefont{Greenlaw et~al.}(1995)\citenamefont{Greenlaw, Hoover,
  and Ruzzo}}]{GrHoRu}
\bibinfo{author}{\bibfnamefont{R.}~\bibnamefont{Greenlaw}},
  \bibinfo{author}{\bibfnamefont{H.~J.} \bibnamefont{Hoover}},
  \bibnamefont{and} \bibinfo{author}{\bibfnamefont{W.~L.} \bibnamefont{Ruzzo}},
  \emph{\bibinfo{title}{Limits to Parallel Computation: {$P$}-completeness
  Theory}} (\bibinfo{publisher}{Oxford University Press},
  \bibinfo{year}{1995}).

\bibitem[{\citenamefont{Hajra and Sen}(2004)}]{HaSe04}
\bibinfo{author}{\bibfnamefont{K.}~\bibnamefont{Hajra}} \bibnamefont{and}
  \bibinfo{author}{\bibfnamefont{P.}~\bibnamefont{Sen}} (\bibinfo{year}{2004}),
  \bibinfo{note}{cond-mat/0406332}.

\bibitem[{\citenamefont{Onody and de~Castro}(2004)}]{OnCa04}
\bibinfo{author}{\bibfnamefont{R.}~\bibnamefont{Onody}} \bibnamefont{and}
  \bibinfo{author}{\bibfnamefont{P.}~\bibnamefont{de~Castro}}
  (\bibinfo{year}{2004}), \bibinfo{note}{cond-mat/0402315}.

\end{thebibliography}
\end{document}